\documentclass[aps,prl,twocolumn,tightenlines,showpacs,aps,amsmath,amssymb,nofootinbib]{revtex4-1}

\usepackage{graphicx}
\usepackage{amsmath,amssymb}
\usepackage{bm}
\usepackage{epsfig}
\usepackage{color}              
\usepackage{comment}
\usepackage{hyperref}

\def\be{\begin{equation}}
\def\ee{\end{equation}}
\def\ba{\begin{eqnarray}}
\def\ea{\end{eqnarray}}
\def\bal{\begin{align}}
\def\eal{\end{align}}
\def\bald{\begin{aligned}}
\def\eald{\end{aligned}}

\newcommand{\per}{\, .}
\newcommand{\com}{\, ,}
\newcommand{\eref}[1]{Eq.~(\ref{#1})}

\newcommand{\muy}{\mu_{\rm \scriptscriptstyle {\cal Y}}}
\newcommand{\muw}{\mu_{\rm \scriptscriptstyle {\cal W}}}
\newcommand{\uwuy}{\rm{U(1)}_{\scriptscriptstyle {\cal W}} \times \rm{U(1)}_{\scriptscriptstyle {\cal Y}}}
\newcommand{\uzua}{\rm{U(1)}_{\scriptscriptstyle {\cal Z}} \times \rm{U(1)}_{\scriptscriptstyle {\cal A}}}
\newcommand{\ua}{\rm{U(1)}_{\scriptscriptstyle {\cal A}}}
\newcommand{\uz}{\rm{U(1)}_{\scriptscriptstyle {\cal Z}}}

\begin{document}

\title{Superconductivity at Any Temperature
}

\date{\today}

\author{Mohamed M. Anber}
\email[]{mohamed.anber@epfl.ch}
\author{Yannis Burnier}
\email[]{yannis.burnier@epfl.ch}
\author{Eray Sabancilar} 
\email[]{eray.sabancilar@epfl.ch}
\author{Mikhail Shaposhnikov}
\email[]{mikhail.shaposhnikov@epfl.ch}
\affiliation{ 
Institut de Th\'eorie des Ph\'enom\`enes Physiques, Ecole Polytechnique F\'ed\'erale de Lausanne, 
CH-1015 Lausanne, Switzerland.
}

\begin{abstract}
We construct a 2+1 dimensional model that sustains superconductivity at all temperatures. This is achieved by introducing a Chern Simons mixing term between two Abelian gauge fields ${\cal A}$ and ${\cal Z}$. The superfluid is described by a complex scalar charged under ${\cal Z}$, whereas a sufficiently strong magnetic field of ${\cal A}$ forces the superconducting condensate to form at all temperatures. In fact, at finite temperature, the theory exhibits Berezinsky-Kosterlitz-Thouless phase transition due to proliferation of topological vortices admitted by our construction. However, the critical temperature is proportional to the magnetic field of $
{\cal A}$, and thus, the phase transition can be postponed to high temperatures by increasing the strength of the magnetic field. This model can be a step towards realizing the long sought room temperature superconductivity.

\end{abstract}
\pacs{11.15.Wx, 11.15.Yc, 11.25.-w
}

\maketitle

\emph{Introduction.}---Superconductivity is one of the most fascinating phenomena in nature that has attracted the attention of both theorists and experimentalists since its discovery in 1911. Superconductors exhibit the so called Meissner effect \cite{meissner33}, namely, the expulsion of external magnetic field lines. It was London brothers who first gave a phenomenological understanding of the Meissner effect \cite{London35}. A breakthrough idea was developed by Landau and Ginzburg who provided a framework to describe superconductivity using mean field theory approach \cite{Ginzburg:1950sr}. In their macroscopic theory of superconductivity, the superfluid is described by a complex scalar field whose expectation value is the order parameter that distinguishes between the superconducting and normal phases. The microscopic structure of the condensate was explained by Bardeen, Cooper and Schrieffer \cite{Bardeen:1957mv} as the pairing of electrons via phonon interactions. Yet another breakthrough came during the 1980s, when it was discovered that certain materials become superconductors at relatively high temperatures, $T\sim 90-130$ K. Since then, it has remained a true challenge to achieve superconductivity at higher temperatures, ultimately all the way up to room temperature. In fact, constructing a model that has a superconducting phase at high temperatures can be a step towards realizing this quest.

In this letter, we report on a 2+1 dimensional $\uzua$ Chern Simons theory that sustains a superconducting phase up to arbitrarily high temperatures. The condensate in this model is described by a complex scalar field that is charged under $\uz$, whose magnetic field exhibits the Meissner effect. Unlike the Abelian Higgs model, whose local U(1) symmetry gets restored at finite temperature, our $\uz$ remains broken at all temperatures. This is achieved by introducing a $\ua$ magnetic field, $B_{\cal A}$, which forces the scalar field to have a non-zero vacuum expectation value. At zero temperature our theory allows a constant magnetic field solution, $B_{\cal A}$, since the $\ua$ symmetry is unbroken. Quantum corrections cannot spoil this symmetry thanks to the Coleman-Hill theorem \cite{Coleman:1985zi,Laine:1999zi}. In fact, this theory also admits topological vortices that are charged under $\ua$, hence they exhibit long range interactions \cite{Anber:2015kxa,Bvortex}. At zero temperature, these vortices do not alter the vacuum structure of the theory as it is costly to excite them. However, at high temperatures the system exhibits a Berezinsky-Kosterlitz-Thouless (BKT) phase transition \cite{Berezinsky:1970fr,Kosterlitz:1973xp} as the vortices proliferate and lead to the breaking of $\ua$. This happens at critical temperature $T_c \propto B_{\cal A}$, and hence, the BKT transition can be pushed to arbitrarily high temperatures by increasing $B_{\cal A}$. Therefore, this theory can support a superconducting phase at all temperatures provided that the external magnetic field $B_{\cal A}$ is large enough. 

\emph{$\uwuy$ Theory.}---We first consider a  topologically massive $\uwuy$ Chern-Simons Higgs theory. The action for two Abelian gauge fields $\mathcal{Y}_\mu$ and $\mathcal{W}_\mu$ coupled to a complex scalar field\footnote{This complex scalar can emerge as an effective description of more fundamental physics.}, $\varphi=(\varphi_1+i\varphi_2)/\sqrt{2}$, reads:
\ba \bald \label{action1}
S &= \int d^3x \biggl [-\frac{1}{4} \mathcal{Y}_{\mu\nu} \mathcal{Y}^{\mu \nu} -\frac{1}{4} \mathcal{W}_{\mu\nu} \mathcal{W}^{\mu \nu}+  \muy \epsilon^{\mu \nu \alpha} \mathcal{Y}_{\mu\nu} \mathcal{Y}_{\alpha} \\ 
& \hskip 1.5cm - \muw \epsilon^{\mu \nu \alpha} \mathcal{W}_{\mu\nu} \mathcal{W}_{\alpha} + |D_\mu \varphi|^{2}- V(\varphi,\varphi^*)\biggr ] \com~~~~
\eald
\ea
where 
\be\label{V phi}
V(\varphi,\varphi^*)=m^2 \varphi\varphi^* +\frac{\lambda}{4} |\varphi\varphi^* |^{2} \com
\ee
and $\mathcal{Y}_{\mu \nu} = \partial_\mu \mathcal{Y}_\nu - \partial_\nu \mathcal{Y}_\mu$, $\mathcal{W}_{\mu \nu} = \partial_\mu \mathcal{W}_\nu - \partial_\nu \mathcal{W}_\mu$, $D_\mu = \partial_\mu -i g_1 \mathcal{Y}_\mu - ig_2 \mathcal{W}_\mu$, and the mass square parameter $m^2$ can be taken to be positive or negative. For simplicity, we take $\lambda\ll|m|$, which will not affect the generality of our results.  The Chern Simons coefficients $\muy$, $\muw$, the mass parameter $m$, and the Higgs self coupling $\lambda$ have mass dimension $M$, while the coupling constants $g_1$ and $g_2$ have mass dimension $M^{1/2}$. We use $\epsilon^{012} =1$, the metric $\eta_{\mu\nu} = {\rm diag}(1, -1,-1)$ and natural units, $c=1$, $\hbar =1$, $k_{B} =1$ in what follows. 

\begin{figure}[t]
\begin{center}
\includegraphics[width=68mm]{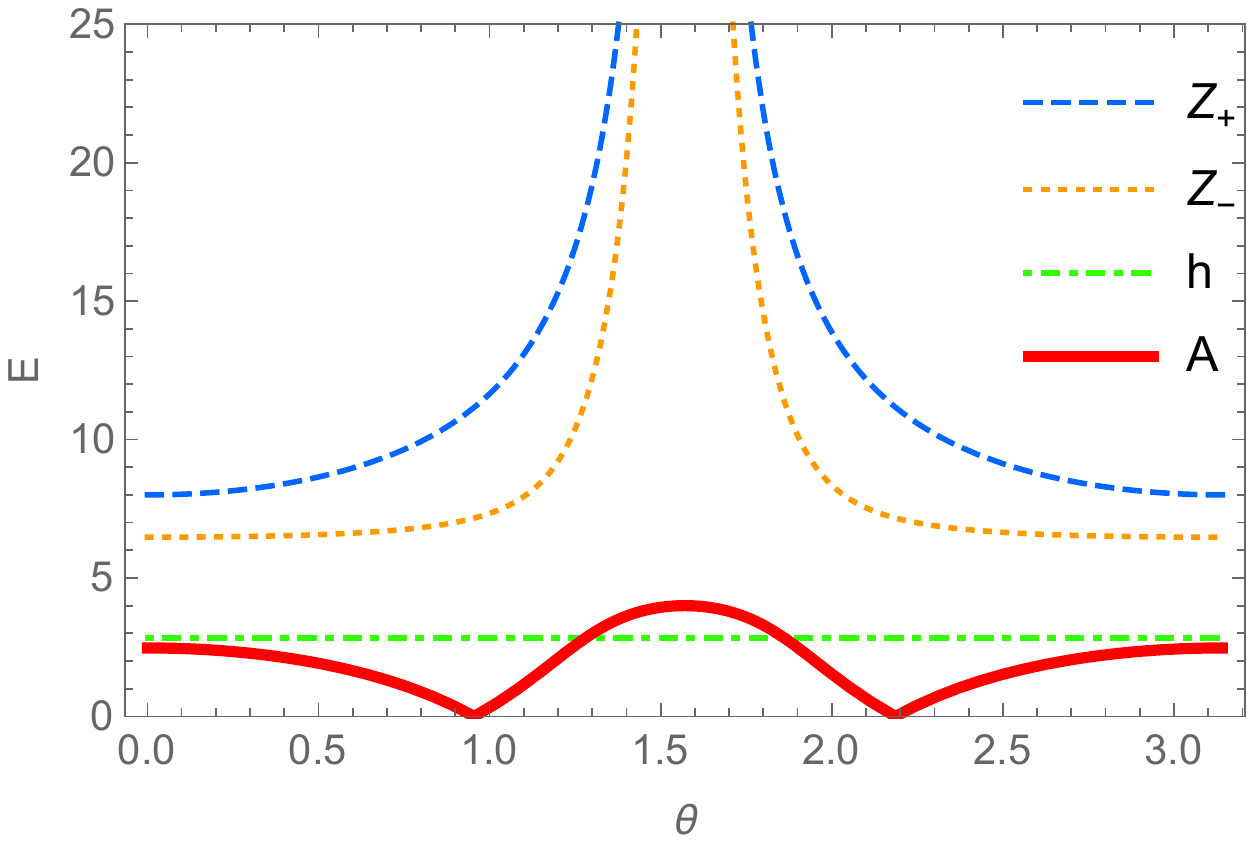}
\vspace{-10pt}
\caption{Masses of the excitations in the $\uwuy$ theory as a function of $\theta$ in the broken phase, $m^2<0$. When $\muy=\muw \tan^2\theta$, the massless mode, ${\cal A}$, appears, and one recovers the $\uzua$ theory described by the action (\ref{action}).}
\vspace{-25pt}
\label{Fig:modes theta}
\end{center}
\end{figure}

The particle spectrum can be easily found by studying the perturbations of the action (\ref{action1}). In the symmetric case, $m^2>0$, we analyze the perturbations
about the Lorentz preserving vacuum $\langle {\cal Y}^\mu \rangle=\langle {\cal W}^\mu \rangle=\langle \varphi \rangle=0$ to obtain the mass spectrum:
\ba\label{modes sym}
\bald
m_{\varphi_1} = m_{\varphi_2}= m \com ~~~m_{\rm \scriptscriptstyle Y} = 4 \muy \com ~~~~m_{\rm \scriptscriptstyle W} = 4 \muw\per
\eald
\ea
A similar analysis can be performed in the broken case, $m^2<0$, where the Higgs field gets a vacuum expectation value $\varphi_0=  \sqrt{\frac{2}{\lambda}}|m|$, $\langle{\cal Y}^\mu\rangle=\langle{\cal W}^\mu\rangle=0$. In this case, it can be shown explicitly that in general all the excitations are again massive.
The explicit expressions are cumbersome combinations of the parameters and will not be given here in their full generality. Instead, we show in Fig.~\ref{Fig:modes theta} the behavior of the mass spectrum as functions of the coupling constants parametrized via the relation $\tan \theta = g_1/g_2$.
We see in Fig.~\ref{Fig:modes theta} that there is a specific combination of the coupling constants ($\muy=\muw \tan^2\theta$) where one of the excitations becomes massless. In the following, we study this specific case in great details.

\begin{figure}[t]
\begin{center}
\includegraphics[width=70mm]{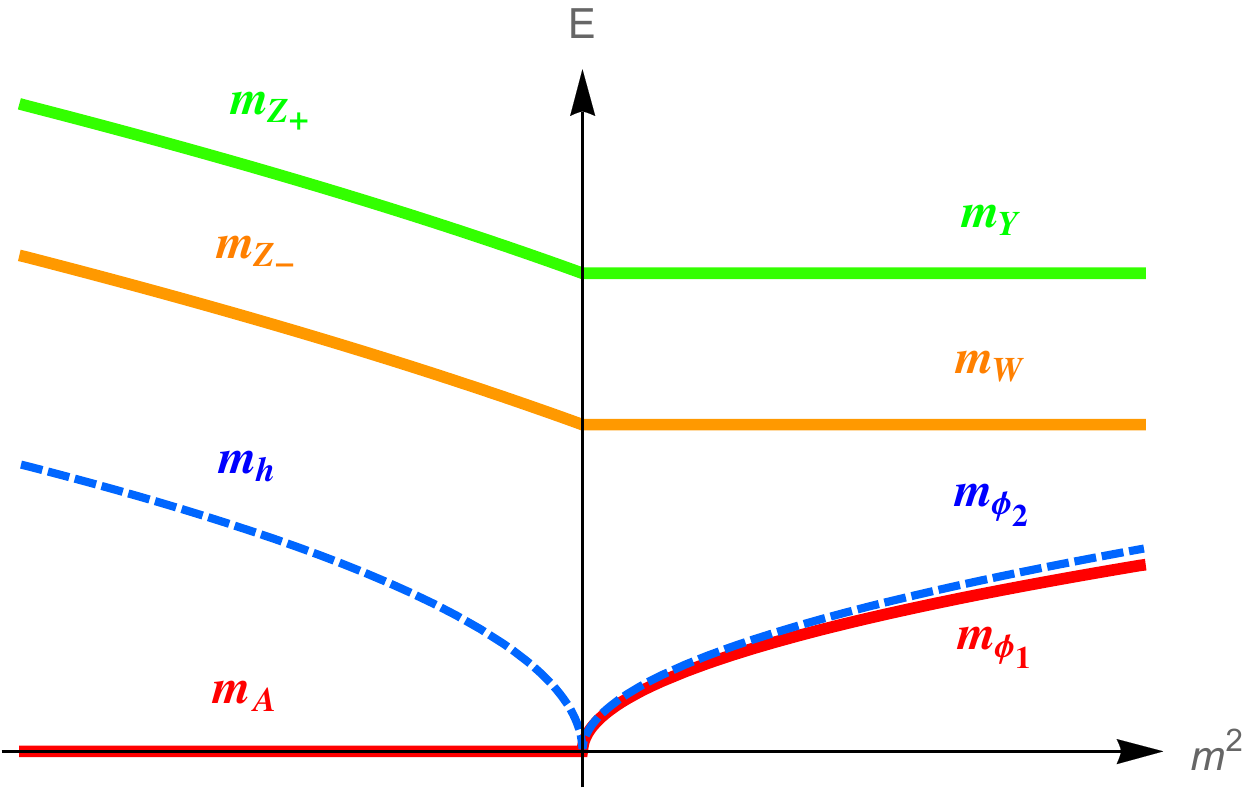}
\vspace{-5pt}
\caption{Masses of the excitations shown in the vacuum, both in the $\uwuy$ and $\uzua$ bases for the case $\muy = \muw \tan^2\theta$, where there is one massless mode ${\cal A}$. (Note that $m_{\varphi_1}=m_{\varphi_2}$, but we artificially separated them for visual clarity.) }
\vspace{-25pt}
\label{Fig:modes}
\end{center}
\end{figure}

\emph{ $\uzua$ Theory.}---When the condition $\muy=\muw \tan^2\theta$ is satisfied, the structure of the spectrum can be understood better in a new basis ${\cal A}_\mu$ and ${\cal Z}_\mu$ defined by $\mathcal{Y}_\mu  = \cos\theta \mathcal{A}_\mu + \sin\theta \mathcal{Z}_\mu$ and $ \mathcal{W}_\mu  = -\sin\theta \mathcal{A}_\mu + \cos\theta \mathcal{Z}_\mu$.
Performing the change of basis and setting $\muy = \muw \tan^2 \theta$, we obtain the $\uzua$ action:
\ba \bald \label{action}
S &= \int d^3x \biggl [-\frac{1}{4} \mathcal{F}_{\mu\nu} \mathcal{F}^{\mu \nu} -\frac{1}{4} \mathcal{Z}_{\mu\nu} \mathcal{Z}^{\mu \nu}+ \mu_1 \epsilon^{\mu \nu \alpha} \mathcal{F}_{\mu\nu} \mathcal{Z}_{\alpha}~~~~~ \\ 
& \hskip 1.5cm + \frac{\mu_2}{2} \epsilon^{\mu \nu \alpha} \mathcal{Z}_{\mu\nu} \mathcal{Z}_{\alpha} + |D_\mu \varphi|^{2}-V(\varphi,\varphi^*)  \biggr ] \per
\eald
\ea
Here $\mathcal{F}_{\mu \nu} = \partial_\mu \mathcal{A}_\nu - \partial_\nu \mathcal{A}_\mu$, $\mathcal{Z}_{\mu \nu} = \partial_\mu \mathcal{Z}_\nu - \partial_\nu \mathcal{Z}_\mu$, $D_\mu = \partial_\mu - i e \mathcal{Z}_\mu$, $V(\varphi,\varphi^*)$ is given by \eref{V phi}, $e = \sqrt{g_1^2 + g_2^2}$, $\mu_1 =  2\muw \tan \theta$ and $\mu_2 = 2 \muw (\tan^{2}\theta-1)$. Without loss of generality we take $\mu_1, \mu_2,e>0$.
The particle spectrum can be studied as in the $\uwuy$ theory and obviously match the previous ones when we set $\muy = \muw \tan^2 \theta$. In the symmetric phase, $m^2>0$, these modes are exactly given by \eref{modes sym}.
In the broken phase, $m^2<0$, we expand $\varphi=\sqrt{\frac{2}{\lambda}}|m|+\frac{h}{\sqrt{2}}$ to find the masses
\ba\label{modes broken}
\bald
m_{h} &= \sqrt{2}|m| \com ~~~~~~~m_{{\cal A}} =0 \com\\
m_{{\cal Z}} &=\pm \mu_2 + \sqrt{\mu_2^2 + 4\mu_1^2 + 4 |m^2| e^2/\lambda}\per
\eald
\ea
The dependence of the spectrum  on $m^2$, for both cases $m^2>0$ and $m^2<0$, is shown in Fig.~\ref{Fig:modes}. Notice how the massless mode corresponding to the unbroken $\ua$ field ${\cal A}_\mu$ emerges. 
Of course one should wonder if quantum corrections can generate a term of the form $\epsilon^{\mu \nu \alpha} {\cal F}_{\mu\nu} {\cal A}_\alpha$, and hence the ${\cal A}_\mu$ field can have a topological mass. Such corrections are not generated at the one-loop level\footnote{In fact, a one loop Feynman diagram involving two external ${\cal A}_\mu$ fields would require two insertions of the Chern-Simons mixing term $\mu_1 \epsilon^{\mu \nu \alpha} \mathcal{F}_{\mu\nu} \mathcal{Z}_{\alpha}$ and would be proportional to the external momentum squared, hence can not give any contribution to the mass.}. Furthermore, the Coleman-Hill theorem \cite{Coleman:1985zi} guarantees that no further correction will be generated at any higher loop order (see also Ref.~\cite{Laine:1999zi}).

%

\emph{Chern Simons Theory in the Background of a Condensate and Magnetic Field.}---In this section, we shall study the  $\uzua$ Chern-Simons theory in the background of a ${\cal Z}_b^0$-condensate and constant magnetic field $B_{\cal A}$. Throughout this section, we take $m^2>0$. We will uncover peculiarities  that are specific to the model at hand, which enable us to engineer a superconducting phase at all temperatures. To this end, we vary the action (\ref{action}) to obtain the equations of motion: 
\ba \bald \label{field equations}
&\partial_\beta \mathcal{F}^{\beta \sigma} + \mu_1 \epsilon^{\beta \alpha \sigma} \mathcal{Z}_{\beta \alpha} = 0 \com \\
&\partial_\beta \mathcal{Z}^{\beta \sigma} + \mu_1 \epsilon^{\beta \alpha \sigma} \mathcal{F}_{\beta \alpha} + \mu_2 \epsilon^{\beta \alpha \sigma} \mathcal{Z}_{\beta \alpha} +j^\sigma=0 \com \\ 
&D_\beta D^{\beta} \varphi +  (m^2 + \frac{\lambda}{2} \varphi\varphi^*) \varphi=0 \com
\eald 
\ea
where $j^{\sigma} = i e \bigl [ \varphi^{*} D^{\sigma} \varphi - (D^{\sigma} \varphi)^{*} \varphi \bigr ]$.

These equations are satisfied by the simple solution
\begin{eqnarray}\label{simple solution}
{\cal B}_{\cal A}=B\,,\quad {\cal Z}_b^0=-\frac{\mu_1 B}{e^2|\varphi|^2}\,,
\end{eqnarray}
where $B$ is a constant. Substituting ${\cal Z}^\mu \rightarrow {\cal Z}^\mu+\delta^{\mu 0}{\cal Z}^0_b$ into the action (\ref{action}),  we obtain the effective potential
\begin{eqnarray} \label{V eff B vacuum}
V_{\mbox{\scriptsize eff}}(|\varphi|,B)=\frac{\mu_1^2B^2}{e^2|\varphi|^2}+m^2|\varphi|^2+\frac{\lambda}{4}|\varphi|^4\,.
\end{eqnarray}
The first term in $V_{\mbox{\scriptsize eff}}(|\varphi|,B)$ is responsible for the exotic phase of matter that we realize in this model, as we will show momentarily.
The minimum of the  potential $V_{\rm eff}(|\varphi|,B)$ is given by
\be\label{B minima}
|\varphi_0| =  \sqrt{\frac{\mu_1 \left|B\right|}{em}}\left(1-\frac{\lambda}{8m}\frac{\mu_1 |B|}{m^2}+{\cal O}\left(\frac{\lambda}{m}\right)^2\right) \com 
\ee
and the condensate at the minimum is 
\begin{eqnarray}
{\cal Z}_b^0(|\varphi_0|) = -\frac{m}{e} \frac{|B|}{B}\left(1+\frac{\lambda}{4m}\frac{\mu_1 |B|}{m^2}+{\cal O}\left(\frac{\lambda}{m}\right)^2\right) \per~~~~
\end{eqnarray}
Therefore, even when $m^2>0$, the theory has a non-zero vacuum expectation value in the presence of  magnetic field.   We stress that the Lorentz symmetry is broken in this vacuum due to the presence of the ${\cal Z}_b^0(|\varphi_0|)$-condensate. We also note that to leading order in $\lambda/m$ the magnitude of the ${\cal Z}_b^0(|\varphi_0|)$-condensate is independent of the magnetic field.
As can be seen from \eref{V eff B vacuum}, regardless of whether $m^2<0$ or $m^2>0$, the $\uz$ symmetry is always broken by the presence of a constant magnetic field $B_{\cal A} = B$.
Similar to the $\uwuy$ theory, the particle spectrum can be found by studying the perturbations of the action (\ref{action}) in the background (\ref{simple solution}): ${\cal A}^{\mu} \simeq  A^\mu \com ~~ {\cal Z}^{\mu} \simeq {\cal Z}_b^{\mu}+Z^\mu \com ~~ \varphi \simeq \varphi_0 + h/\sqrt{2}$.
 The mass of the excitations in this background are ($m^2>0$)
\ba\label{modes broken generic}
\bald
m_{h} &= \sqrt{3\lambda |\varphi_0|^2/2 +m^2 + 3 e^2 {{\cal Z}_b^0}^2} \com ~~~~m_{\cal A} =0 \com~~~~\\
m_{\cal Z} &= \pm \mu_2 + \sqrt{\mu_2^2 + 4\mu_1^2 + 2e^2 |\varphi_0|^2 }\per
\eald
\ea
Exactly as in \eref{modes broken}, we find that the $\ua$ mode is massless. The massless  $\ua$ is consistent with our assumption that the vacuum can support a constant magnetic field $B_{\cal A}$. In addition,  the massless  $\ua$ plays a pivotal role in the nonperturbative physics, as we discuss in the next section.

It is surprising to realize a system where a small external magnetic field breaks the symmetry even when $m^2>0$. One may wonder whether the symmetry can be restored as we heat up the system. As we shall briefly see, the symmetry remains broken even at high temperatures provided that the external magnetic field is large enough.

\emph{Nonperturbative Effects.}---The $\uzua$ Chern-Simons theory described by \eref{action} with a non-zero vacuum expectation value (either in the case $m^2<0$, or in the case $m^2>0$ with constant magnetic field) admits topological vortex solutions which carry winding numbers corresponding to the first homotopy group of the vacuum manifold. We show their existence by analytic and numerical means in Ref.~\cite{Anber:2015kxa,Bvortex}.  These vortices are charged under the unbroken $\ua$. Moreover, in the background magnetic field,  they act as diamagnetic material since $\ua$ becomes topologically massive near the core region. In addition,  a vortex does not transform into anti-vortex since the vortex solution breaks both $C$ and $P$ symmetries while preserving $CP$, see Ref.~\cite{Bvortex} for more details. In fact, a vortex-anti-vortex pair exhibits logarithmic confinement as charged particles would do in $2+1$ dimensions.  A pair of the lowest winding vortices separated a distance $R$ (a charge neutral combination) yields a finite energy configuration \cite{Anber:2015kxa,Bvortex}:
\begin{eqnarray}
E=\frac{8\pi|\varphi_0|^2\mu_1^2}{e^2|\varphi_0|^2+2\mu_1^2}\log\frac{R}{r_{\rm c}}+2E_{\rm c}\,,
\end{eqnarray}
where $r_{\rm c}$ is the vortex core radius, and  $E_{\rm c}\cong 2\pi\varphi_0^2$ is the core energy.

At zero temperature, it is very expensive to excite the vortices and therefore they do not alter the structure of the vacuum. However, they become the main players at finite temperature, as we discuss below.

\emph{Superconductivity at Zero Temperature.}---The $\uzua$ Chern-Simons theory, with $m^2<0$ or $m^2>0$ (in the later case we have to turn on a constant magnetic field), has the correct ingredients to be a good candidate for an effective field theory of superconductivity in $2+1$ D. The theory admits a superfluid phase which is characterized by a non-zero vacuum expectation value $\varphi_0$. When the phase of $\varphi$ is gauged, the would be Goldstone bosons are eaten by the $\uz$ field. Then, the ${\cal Z}_\mu$ gauge boson acquires a mass and the Meissner effect sets in characterizing a superconducting phase. Therefore, the theory can conduct $\uz$ currents, $j^\mu_{\cal Z}=ie \varphi^*\partial^\mu \phi-ie\varphi^*\partial^\mu\varphi+2e^2|\varphi|^2{\cal Z}^\mu$, with no dissipation.  On the other hand, the massless $\ua$ does not couple directly to $\varphi$, and hence it is a ``dark sector" of the theory.

What we have said so far about superconductivity is a standard material that can be realized in a simple U(1) Abelian Higgs model. So what is new about our model? It is well known that at finite temperature we lose superconductivity in the Abelian Higgs model due to symmetry restoration. On the contrary, our model provides an example where this restoration does not happen, or at least can be postponed until extremely high temperatures\footnote{The non-restoration of  symmetries at high temperatures may also occur in gauge theories with extended Higgs sector with a certain choice of gauge and scalar coupling constants \cite{Mohapatra:1979vr}.}. This is the topic of our next section.

\emph{Superconductivity at Finite Temperature.}---It is crucial for our construction that the $\ua$ gauge field remains massless. As was previously stressed, at zero temperature the Coleman-Hill theorem forbids the generation of $\epsilon^{\mu \nu \alpha} {\cal F}_{\mu\nu} {\cal A}_\alpha$ term. However, at finite temperature, the Coleman-Hill theorem does not apply, and thus, one has to check explicitly whether such a term will be generated. We verified that this term is absent in one and two-loop calculations. 

At finite temperature, one integrates all the heavy particles in the system, the Higgs, the ${\cal Z}$-boson, as well as their Kaluza-Klein excitations.  Up to the leading order in $\lambda/|m|$, we obtain the effective potential:
\ba\label{V_eff T B}
\bald
V_{\mbox{\scriptsize eff}}({\footnotesize \text{$|\varphi|,B,T$}}) &=  \frac{\mu_1^2 B^2}{e^2|\varphi|^2} +m^2 |\varphi|^2 + \frac{\lambda}{4}|\varphi|^4 +\Pi({\footnotesize \text{$|\varphi|,B,T$}})\,,\\
\Pi(|\varphi|,B,T)&=\sum_{a=1,2}\frac{T{\cal C}_a}{2}\int \frac{d^2k}{(2\pi)^2}\log\left [1-e^{-\omega_a/T}\right] \,.
\eald
\ea
$\omega_a=\omega_a(k,|\phi|,B)$ are the dispersion relations for the Higgs and ${\cal Z}$-boson (including the ghost), and ${\cal C}_a$ are multiplicity factors\footnote{The dispersion relations in the presence of $B_{\cal A}$ magnetic field are cumbersome, and we do not give them here.}. For $B=0$ and $m^2<0$ (in the superconducting phase), we find that the high $T$ limit of $\Pi(|\varphi|,B,T)$ behaves as $(2e^2 +\lambda) |\varphi|^2~ T \log\left[\frac{T}{m_{\cal Z}} +{\rm const} \right]$. Thus, at $T \sim m_{\cal Z}$ the $\uz$  symmetry is restored and  superconductivity is lost, exactly like in the case of Abelian Higgs model. The situation changes dramatically in the presence of $B_{\cal A}$ magnetic field, thanks to the first term in (\ref{V_eff T B}). We checked numerically that $\Pi(|\varphi|,B,T)$ can not compete with $\frac{\mu_1^2 B^2}{e^2|\varphi|^2}$ at any value of $B$ and $T$. Thus, the presence of $B_{\cal A}$ magnetic field, even a small $B$, will ensure that the theory has a non zero vacuum expectation value at all temperatures, hence broken\footnote{ Also, we do not expect the situation can change to any loop order since the classical term $\frac{\mu_1^2 B^2}{e^2|\varphi|^2}$ is non-analytic in $|\varphi|$ and hence can not be beaten by perturbative effects.} $\uz$. 

So far, our finite-temperature analysis was based on perturbation theory. However, the $\uzua$ Chern-Simons theory admits nonperturbative objects; namely these are the long-range vortices mentioned above and discussed in  our accompanying works \cite{Anber:2015kxa,Bvortex}. Although it is very expensive to excite these objects at zero temperature, and hence they do not alter the vacuum, at finite temperature they are only suppressed by the Boltzmann factor (fugacity) $e^{-E_c/T}$, where $E_c\cong 2 \pi\varphi_0^2$ is the vortex core energy. Therefore, at non-zero temperature the absence or presence of these objects is determined by a competition between the energy, $E$, and entropy, $S$, of the system. The free energy, ${\cal F}=E-TS$, of ${\cal N}$ vortices and ${\cal N}$ anti-vortices in a system of a linear size $L$ is ${\cal F}={\cal N}\left[2\eta\log\frac{R}{r_c}+2E_c \right]-2{\cal N}T\log\frac{L^2}{r_c^2}+2{\cal N}T \log {\cal N} = L^2\frac{2}{R^2}(\eta-2T)\log\left[\frac{R}{r_c} e^{E_c/(\eta-2T)}\right]$,
where $\eta=\frac{4\pi|\varphi_0|^2\mu_1^2}{e^2|\varphi_0|^2+2\mu_1^2}$ is half the interaction energy between a vortex-anti-vortex pair, and $R=L/\sqrt{{\cal N}}$ is the mean separation between the vortices\footnote{Remember that the vortex-anti-vortex interaction energy is solely due to the unbroken $\ua$.}. At low temperatures,  $T<\eta/2$, the free energy  is minimized at $R\rightarrow \infty$, and therefore, it is very expensive to excite the vortices. However, at high temperatures, $T>\eta/2$, the free energy is minimized at  $R\sim e^{E_c/(2T-\eta)}~r_c $ and the vortices proliferate in the vacuum. Hence, a phase transition happens at the critical temperature
\begin{eqnarray}\label{KT TEMP}
T_c=\frac{2\pi|\varphi_0|^2\mu_1^2}{e^2|\varphi_0|^2+2\mu_1^2}\,,
\end{eqnarray}
see Ref.~\cite{Bvortex} for a more rigorous derivation. In fact, this is the celebrated Berezinsky-Kosterlitz-Thouless transition \cite{Berezinsky:1970fr,Kosterlitz:1973xp}. At temperatures $T\geq T_c$ electric fields are Debye screened and the ground state of the theory becomes more complicated.

In the absence of $B_{\cal A}$ magnetic field, $T_c$ could happen either before or after the restoration of the $\uz$ symmetry depending on the parameters of the theory.   At a non-zero $B_{\cal A}$, we substitute \eref{B minima} into \eref{KT TEMP} to obtain $T_c=\frac{2\pi\mu_1^2B_{\cal A}}{e^2B_{\cal A}+2em\mu_1}$.
Now, if we take $\mu_1\gg eB_{\cal A}/(2m)$, we find $T_c\cong\pi\mu_1 B_{\cal A}/(em)$. Thus, by increasing the external $B_{\cal A}$ magnetic field, keeping the hierarchy $\mu_1\gg eB_{\cal A}/(2m)$, we can ensure that $\ua$ remains unbroken as we increase the temperature. Also, since $\uz$ remains broken, we can engineer a superconducting phase  at very high temperatures\footnote{A similar phenomenon where the magnetic field can suppress phase transition was also noticed in Ref.~\cite{Anber:2013tra}.}. Notice that in the $\mu_1\gg eB_{\cal A}/(2m)$ limit, the kinetic term of the $\ua$ field can be ignored compared to the $\mu_1 \epsilon^{\mu \nu \alpha} \mathcal{F}_{\mu\nu} \mathcal{Z}_{\alpha}$ term. Therefore, one only needs a non-dynamical $\ua$ to achieve superconductivity at arbitrarily high temperatures. In fact, such non-dynamical fields can be synthesized in condensed matter experiments (see, e.g., \cite{RevModPhys.83.1523}).

 U(1)$\times $U(1) Chern Simons models have been considered extensively in the condensed matter literature, see e.g., \cite{PhysRevB.42.8133,PhysRevB.42.8145,PhysRevB.46.2290,PhysRevB.80.205319}. In this letter, we reported on a new phase of matter in $\uzua$ Chern-Simons theory with a complex scalar and argued that this theory can sustain a superconducting phase to arbitrarily high temperatures. Synthesizing a material that is described by action (\ref{action}) could answer the quest for building room temperature superconductors.  

We would like to thank Vadim Cheianov for useful discussions. This work was supported by the Swiss National Science Foundation. Y.B. is supported by the grant PZ00P2-142524.

\bibliographystyle{apsrev4-1}
\bibliography{cssymmetric_refs}

\end{document}